# A numerical solution of the inverse problem in classical celestial mechanics, with application to Mercury's motion

**M. Arminjon**

Laboratoire "Sols, Solides, Structures"
[CNRS / Université Joseph Fourier / Institut National Polytechnique de Grenoble]
B.P. 53, 38041 Grenoble cedex 9, France.
E-mail: arminjon@hmg.inpg.fr

**Abstract.** It is attempted to obtain the masses of the celestial bodies, the initial conditions of their motion, and the constant of gravitation, by a global parameter optimization. First, a numerical solution of the *N*-bodies problem for mass points is described and its high accuracy is verified. The osculating elements are also accurately computed. This solution is implemented in the Gauss iterative algorithm for solving nonlinear least-squares problems. This algorithm is summarized and its efficiency for the inverse problem in celestial mechanics is checked on a 3-bodies problem. Then it is used to assess the accuracy to which a Newtonian calculation may reproduce the DE403 ephemeris, that involves general-relativistic corrections. The parameter optimization allows to reduce the norm and angular differences between the Newtonian calculation and DE403 by a factor 10 (Mercury, Pluto) to 100 (Venus). The maximum angular difference on the heliocentric positions of Mercury is *ca.* 220" per century before the optimization, and *ca.* 20" after it. The latter is still far above the observational accuracy. On the other hand, Mercury's longitude of the perihelion is not affected by the optimization: it keeps the linear advance of 43" per century.

**Key words:** Mercury's perihelion. Parameter optimization. Mechanics of point masses.

## 1. Introduction

In order to predict the motion of celestial bodies, one has to know the values of the basic parameters of the *N*-bodies problem: masses, initial conditions, Newton's constant, and (*e.g.* for the Earth-Moon system) some higher-order multipoles of the mass distribution. In the past, these values were adjusted from the consideration of partial systems: for instance, the masses of the planets were obtained by applying Kepler's third law for the two-bodies problem, successively to the system constituted by the Sun and one planet, and to the system of this planet and a small satellite of it [1]. Due to the progress in the computer capacities, it has become possible to proceed to more global adjustments. Thus, the construction of ephemerides, based on a numerical [2] or analytical [3-4] solution of the *N*-bodies problem, includes least-squares fittings [2,5]. Among the solved-for parameters, many are relevant to data processing: *e.g.* catalogue drift parameters, phase corrections, *etc.,* but the initial conditions are also solved for at this stage [2,5]. However, the masses usually are not solved for, instead they are still obtained from the consideration of partial systems, with the "satellite" being often replaced by a spacecraft in close approach.

Moreover, the construction of ephemerides includes corrections based on general relativity (GR) [2,6]. The optimal parameters depend on the precise model that is used. In particular, they should be reevaluated if one tests a new theory of gravitation [7]. An optimization program has been built in the latter context and, in order to check the algorithm, it has first been tested with the purely Newtonian equations of motion. The aim of this paper is to discuss this Newtonian parameter optimization. As an application, it is investigated in what measure one may reproduce, with a purely Newtonian calculation, the predictions of an ephemeris that includes corrections based on GR. In the context of testing theories of gravitation, the time interval to be considered is of the order of one century, because in the last century the accuracy of the observations has grown so much as to make earlier observations almost useless. Old observations (*e.g.,* of eclipses) may be used for ultimate checks, which shall not be investigated in this paper.

Section 2 describes the numerical solution of the Newtonian *N*-bodies problem and its test. The algorithm used in our global parameter optimization is discussed in Sect. 3, that also shows a test of the efficiency of the algorithm. The application to the fitting of a "relativistic" ephemeris by a Newtonian calculation is presented and discussed in Sect. 4. Our conclusions are given in Sect. 5. We emphasize that Sect. 2 has no claim of scientific novelty, and that the only original features in Sect. 3 may be Eq. (8), plus the use of directional minimization, which indeed seems to be new in the context of celestial mechanics (though of course not in optimization in general). Yet it has been found useful to write these two Sections rather in detail, firstly for the convenience of the readers of this Journal, also because Sect. 2 "qualifies" the present calculations, and because it is difficult to find explicit descriptions of the adjustment algorithms used in celestial mechanics, as it is given in Sect. 3 for the algorithm used here.

## 2. Numerical solution of the Newtonian *N*-bodies problem, and its test

The system is constituted by $N$ extended bodies $B_1, ..., B_N$, subjected to Newtonian gravitation. Assuming that the maximum of the ratio *L/R*, with *L* the size of a body and *R* its distance from another body, is a small parameter $\eta$, and neglecting $\eta^2$ with respect to unity, we may consider these bodies as point masses (*cf.* Fock [8], §71; for nearly spherical bodies, the error is smaller; see the end of this Section for one correction to this assumption). Using Cartesian coordinates in inertial directions, centered at the mass center of the $N^{th}$ body (here the Sun), the equations of motion take the usual form

$$\ddot{\mathbf{r}}_i = -G(m_N + m_i)\frac{\mathbf{r}_i}{R_i^3} + \sum_{j=1(j\neq i)}^{N-1} Gm_j\left(\frac{\mathbf{r}_j - \mathbf{r}_i}{\Delta_{ij}^3} - \frac{\mathbf{r}_j}{R_j^3}\right), \qquad (i = 1, ..., N-1) \quad (1)$$

(see *e.g.* Le Guyader [9]), in which $m_i$ and $r_i$ are the mass of body (*i*) and the heliocentric radius vector of its mass center, $R_i = \|r_i\|$, $\Delta_{ij} = \|r_i - r_j\|$, $G$ is Newton's gravitational constant, and an upper dot means time derivative. We use the International Astronomic Units (IAU), thus in particular $m_N = 1$ and $G = k^2$ with $k$ the Gauss constant. To solve Eq. (1) with given initial conditions $r_i(t_0) = r_{i0}$ and $\dot{r}_i(t_0) = u_{i0}$, we set

$$\mathbf{y} = (\mathbf{r}_1, ..., \mathbf{r}_{N-1}, \dot{r}_1, ..., \dot{r}_{N-1}) \quad (2)$$

and

$$f(\mathbf{y}) = (\dot{r}_1, ..., \dot{r}_{N-1}, \ddot{r}_1, ..., \ddot{r}_{N-1}), \quad (3)$$

where each vector $\ddot{r}_i$ is given by Eq. (1). We then have to solve the first-order differential system $\dot{y} = f(y)$ with initial condition $y(t_0) = y_0$. This is done by using the Matlab routine ODE113, a variable order Adams-Bashforth-Moulton solver. This solver adjusts the actual integration step in order to reach two numerical tolerances "RelTol" and "AbsTol", assigned by the user. The program is implemented in double precision on personal computers.

In order to adjust the numerical tolerances and to test the accuracy of the numerical integration, the program was run for the system (1) obtained with the Sun, Mercury and Jupiter, asking back-and-forth time integrations all starting at Julian Day JD 2451600.5 (26/02/2000 0H00), and first with JD 2447600.5 as the oldest date, thus 8000 days. The optimal tolerances (*i.e.*, those leading to the smallest integration errors) were thus found to be close to RelTol = $5 \times 10^{-13}$ and AbsTol = $10^{-15}$. The maximum integration error (the difference between the initial value and the value after the back-and-forth time integration) is then $\delta x_{\text{Mercury}} = 4.69 \times 10^{-10}$ au (one *astronomical unit* (au) = 149597870 km [10, p. 22]). With JD 2411600.5 as the oldest date, thus 80000 days (some 219 years), and with the same tolerances, the maximum error is $\delta x_{\text{Mercury}} = 3.35 \times 10^{-8}$ au. The maximum integration error seems conditioned by the quickest planet (Mercury) and does not increase much as more planets are taken into consideration (*e.g.* it is still $\delta x_{\text{Mercury}}$, now equal to $-1.68 \times 10^{-9}$ au, when the five innermost planets are taken over the same period of 8000 days as above). When Mercury is taken in consideration, the mean of the actual integration step is about 0.5 day. This corresponds well with the results of Schubart & Stumpff [11].

To check the results given by the program, they were compared with the ones published by Le Guyader [9]. Thus, the initial conditions at JD 2451600.5 and the masses of the nine planets were those used by Le Guyader [9: Table 1 for the masses and Table 3 for the initial conditions]. Our final states (referred to the dynamical ecliptic and equinox J2000) at JD 2411600.5 were compared with those given in Table 4 of Le Guyader, that correspond to the same date. This comparison showed that, after 40000 days, our program gives position coordinates differing by $1.6 \times 10^{-7}$ au (Mercury) to $5 \times 10^{-9}$ au (Uranus to Pluto) from those found by Le Guyader, while the velocity components differ by $10^{-8}$ au/d (Mercury) to $2 \times 10^{-13}$ au/d (Pluto). We may conclude that the numerical accuracy of our program is enough for a study over a period of the order of the century, in particular it is largely enough for the discussion, in Sect. 4, of the differences due to different physical models (*cf.* Table 3 below). For much longer periods, one would need either to go from double to quadruple precision, or to use a more sophisticated solution method, such as that proposed by Le Guyader [9]. Due to its greater computation time, the latter method was less appropriate for a parameter optimization, because this involves solving the equations of motion many times.

The calculation of the *osculating elements* was also tested: if, at a given time, one leaves aside all contributions to the acceleration of a given planet (*i*), apart from the acceleration valid for the Newtonian 2-bodies problem with the Sun and that planet [*cf.* Eq. (1)], then the motion is the "osculating" Keplerian motion, determined by the current values of the heliocentric position and velocity vectors $\boldsymbol{r}$ and $\dot{\boldsymbol{r}}$, and by the attraction constant of the 2-bodies problem,

$$\mu = G(m_N + m_i). \tag{4}$$

A routine was built to compute the classical elements of the osculating motion [12] (semi-major axis $a$, eccentricity $e$, inclination $i$, longitude of node $\Omega$, argument of perihelion $\omega$, time at perihelion $\tau$), as well as the energy $H = \dot{\boldsymbol{r}}^2 / 2 - \mu / \|\boldsymbol{r}\|$ and the "standard" set of elements: $a$, the mean longitude $\lambda$, and the quantities $k, h, q, p$ (defined *e.g.* in Le Guyader [9]), from the data $\boldsymbol{r}, \dot{\boldsymbol{r}}$ (and $\mu$). Another routine computes, conversely, $\boldsymbol{r}$ and $\dot{\boldsymbol{r}}$ from the data ($a, \lambda, k, h, q, p$). To test them, the program for the *N*-bodies problem was run with the Sun and a second body, either the Earth-Moon barycenter (EMB) or Venus. It was found that that $a, e, i, \Omega, \varpi \equiv \Omega + \omega$, $\boldsymbol{\sigma} \equiv \boldsymbol{r} \wedge \dot{\boldsymbol{r}}$, and $H$, which should be constant, remain extremely nearly so, and that the values for $a, e, i, \Omega$ and $\varpi$ are close to the mean elements given by Simon *et al.* [10, p. 231]. The inversion was tested by recalculating $\boldsymbol{r}$ and $\dot{\boldsymbol{r}}$ from the calculated ($a, \lambda, k, h, q, p$) and by comparing with the input values of $\boldsymbol{r}$ and $\dot{\boldsymbol{r}}$. For the EMB, for instance, the maximum difference $|r_m - (r_m)_{\text{recalculated}}|$ found for any component $r_m$ of $\boldsymbol{r}$ over 4000 days (stored by steps of 200 days) is less than $2 \times 10^{-14}$ au, and the maximum difference $|\dot{r}_m - (\dot{r}_m)_{\text{recalculated}}|$ is less than $4 \times 10^{-16}$ au/d.

Using the osculating elements, some analytical perturbations of the Moon on the motion of the EMB [13] were taken into account in the program.

## 3. Algorithm for the parameter optimization, and its test

The aim of this work was to simultaneously optimize all parameters of the Newtonian problem for ten mass points (the Sun and the nine major planets), *i.e.* $G = Gm_{\text{Sun}} = Gm_{10}$, the products $Gm_i$ ($i = 1, ..., 9$), and the initial conditions for the heliocentric positions and velocities of the nine planets. In general, we have a set of parameters, $\alpha = (\alpha_k)_{k = 1, ..., K}$, with $K = 64$ in the real problem investigated, and we seek to minimize the residual

$$R(\alpha) \equiv \{\Sigma_j \, [F'_j(\alpha) - D'_j]^2 / \Sigma_j \, w_j^2\}^{1/2}, \quad F'_j(\alpha) \equiv w_j F_j(\alpha), \quad D'_j \equiv w_j D_j, \qquad (5)$$

where $D_j$ ($j = 1, ..., J$) are the input data, $F_j(\alpha)$ is the theoretical prediction for $D_j$, and where the $w_j$'s are weights. In the application, the $D_j$'s were taken to be the heliocentric positions and/or velocities of the planets as given by the ephemeris DE403 [14], at a set of times which may depend on the planet considered. The algorithm for minimizing $R$ is the *Gauss algorithm*, based on the linearization of the functions $F'_j(\alpha)$ around the value of $\alpha$ at the current optimization iteration [15]. Thus, defining

$$A_{jk} = \partial F'_j / \partial \alpha_k, \qquad E_j = F'_j(\alpha) - D'_j, \qquad (6)$$

we solve the system

$$\sum_{k=1}^{K} A_{jk} \, \delta\alpha_k = -E_j \qquad (j = 1, ..., J) \qquad (7)$$

in the sense of the least squares. (Here, the calculation of the functions $F_j(\alpha)$ consists in numerically solving the equations of motion with the current values of the parameters $\alpha_k$, hence the derivatives $\partial F_j / \partial \alpha_k$ are computed by finite differences. This is the most time-consuming step.) It can be shown that the theoretical least-squares solution $\delta\alpha$ satisfies

$$\nabla R'(\alpha) . \delta\alpha = -\|\text{Proj}_{\text{Im}(D\mathbf{F}'(\alpha))} \mathbf{E}\|^2 , \qquad (8)$$

where $\mathbf{E} = (E_j)$ and $\text{Proj}_{\text{Im}(D\mathbf{F}'(\alpha))}$ is the orthogonal projection, in the data space $\mathbf{R}^J$, over the range, $\text{Im}(D\mathbf{F}'(\alpha))$, of the linear mapping $D\mathbf{F}'(\alpha)$ tangent to $\mathbf{F}' = (F'_j)$ at $\alpha$, and where

$$R'(\alpha) \equiv \Sigma_j \, [F'_j(\alpha) - D'_j]^2 = [R(\alpha)]^2 \, \Sigma_j \, w_j^2. \qquad (9)$$

Equation (8) means that the direction $\delta\alpha$ is a direction of descent for $R'$ (hence for $R$ as well), unless the projection on the r.h.s. of Eq. (8) is zero. But it is easy to check that this occurs if, and only if, $\alpha$ is a stationary point for $R'$ (and $R$), *i.e.* $\nabla R'(\alpha) = 0$. If the minimization algorithm leads to a such stationary point, this cannot be a local maximum (except if the algorithm were started from that very maximum – this is very unlikely), hence it should be a minimum for $R$, or possibly a saddle point. In any case, the algorithm can then not do better.

In this work, the foregoing traditional Gauss algorithm was augmented with the *directional minimization:* Once the descent direction $\delta\alpha$ has been computed, the residual is minimized in this direction, *i.e.,* one searches for the minimum of the function of one variable,

$$f(\xi) = R(\alpha + \xi \, \delta\alpha) \qquad (0 \leq \xi \leq \xi_{\sup}). \qquad (10)$$

This is done by using the Matlab routine FMIN, based on golden section search and parabolic interpolation. The upper bound $\xi_{sup}$ of the minimization interval was usually set to 1.5. This one-dimensional minimization is stopped when the position $\xi_0$ of the minimum has been found up to a predefined accuracy $\eta$ (usually $\eta = 10^{-8}$ was selected). The global minimization is brought to an end when either the modification of the parameters becomes smaller than a predefined value, $\xi_0 \|\delta\alpha\| < \varepsilon$, or the last direction $\delta\alpha$ fails to provide a lower value for $R$. To the author's knowledge, the directional minimization was not used before for parameter optimization in celestial mechanics, and indeed it is often not very useful in that context: if one starts from a good estimate of the optimal vector, then the optimal value of $\xi$ is close to 1. However, it has been found that, at the following iterations, the optimal value of $\xi$ is usually smaller than 1, and that sometimes one would get no improvement at all over the first iteration if one would impose $\xi = 1$ (though, when the initial estimate is good, the following iterations do not improve much over the first one). More importantly, if the initial estimate is not good enough, then one must have recourse to the directional minimization, for otherwise even the first iteration brings no improvement over the initial estimate. This does happen, for instance, in the test below.

In order to check the efficiency of the algorithm, the input data $D_j$ were taken to be the heliocentric positions and velocities calculated by the $N$-bodies program with the Sun, Jupiter and Saturn (with the initial conditions at JD 2451600.5 and the masses from Le Guyader [9]), for a time span of –4000 days, stored by steps of 200 days. Thus, the input data spanned less than 40% of the duration of one period for Saturn. After that calculation, a random perturbation between 0 and +0.1, with uniform distribution, was imposed to the 15 parameters $\alpha_k$ ($G$, two masses and 12 initial conditions) that determine its result. I.e., each $\alpha_k$ was replaced by $(1+\xi_k)\alpha_k$, with each $\xi_k$ randomly selected between 0 and 0.1 – thus a very severe perturbation. With these perturbed values of the parameters, the initial value of the residual (5), with equal weights, was found to be $R = 1.20$. Then the minimization algorithm was run. After 8 successful iterations, the program stopped with initial conditions differing by at most $7 \times 10^{-12}$ from the true ones, and with masses equal to the true ones up to a relative error smaller than $10^{-9}$. The relative error on $G$ was $10^{-11}$, and the residual was $R = 1.7 \times 10^{-12}$. Thus, in this test, it is known in advance that the parameter optimization has a solution where the residual is zero, and, to the accuracy allowed by the integration errors, the optimization program indeed finds this solution – although the solution is far removed from the initial (perturbed) set of values of the parameters. This severe test may seem rather artificial, for in celestial mechanics one usually has very good initial estimates of the parameters. It is more justified in the context of testing an alternative theory of gravitation, however, since the adequate values of the parameters depend on the gravitational model and might thus change significantly.

**4. Application: a global parameter optimization for the ten major bodies of the solar system**

*4.1 Input data for the calculation of the residual*

To handle each planet on the same footing, the same number of ephemeris data $D_j$ was taken for each, *i.e.* $n = 21$ points, equally spaced in time for a given planet, and at each point the position and velocity vectors were taken, thus 126 input data per planet were taken from the DE403 ephemeris. The total number of input data was hence $J = 1134$. The data of the DE403 ephemeris are in equatorial Cartesian coordinates based on the J2000 reference frame of the International Earth Rotation Service [14]. All sets of data started from $t_0 =$ JD 2451600.5 (26/02/2000 0H00) and went back in time. For Mercury (planet number $i = 1$), the time interval between two points of data was $\delta t = 20$ days, thus the data for Mercury spanned 400 days, that is some 4.5 periods. In order to try to span the same time, *relative* to the period $T_i$ of each planet, the time interval between data for any other planet (*i*) was "initialized" to $(T_i/T_1)\delta t$, then reduced to match two additional constraints: 1) the soonest date was limited to $t_0$ minus one century; and 2) the ephemeris data were stored with an interval of 20 days. The time space between ephemeris data was thus 40 days for Venus, 80 for the EMB, 140 for Mars, 980 for Jupiter, and 1740 for the four outermost planets. Hence the time span of the calculation was actually 34800 days $\approx$ 95.3 years.

Equal weights were taken for the calculation of the residual $R$, thus all $w_j = 1$ in Eq. (5). Finally, the vector of parameters (54 initial conditions, for the date JD 2451600.5, and ten products $Gm_i$), $\alpha = (\alpha_k)_{k = 1, ..., 64}$, was initialized at the DE403 values.

*4.2 Result of the optimization and discussion*

The initial value of the residual was $3.67 \times 10^{-7}$, and the minimum found was $1.54 \times 10^{-8}$, after four iterations. (In the present case, most of the descent was done at the first iteration.) Thus the mean quadratic difference between the Newtonian calculation and the ephemeris (at the selected points) was reduced by a factor of some 24. Tables 1 and 2 show the optimal values found for the initial conditions, and the optimal ratios $m_{Sun}/m_i$ found, respectively.

**Table 1.** Optimal values for the initial conditions (at JD 2451600.5)

Heliocentric positions (au) for Mercury, ..., Pluto
```
-2.50332204746417E-001  +1.87321749063159E-001  +1.26023015179088E-001
+1.74777924711780E-002  -6.62421034024001E-001  -2.99120322376906E-001
-9.09191616400210E-001  +3.59292577761738E-001  +1.55773048023340E-001
+1.20301883272756E+000  +7.27071305012332E-001  +3.00956152960212E-001
+3.73307681521600E+000  +3.05242440090800E+000  +1.21742846649852E+000
+6.16443481381500E+000  +6.36677300992400E+000  +2.36453204691500E+000
+1.45796485723700E+001  -1.23689144167400E+001  -5.62361980789300E+000
+1.69549169042000E+001  -2.28871459674200E+001  -9.78992618667900E+000
-9.70712790906900E+000  -2.80410116011900E+001  -5.82383440514300E+000
```

Heliocentric velocities (au/d) for Mercury, ..., Pluto
```
-2.43880793709909E-002  -1.85022423354218E-002  -7.35381099390148E-003
+2.00854702758845E-002  +8.36545250335911E-004  -8.94788932399633E-004
-7.08584354283626E-003  -1.45563423137304E-002  -6.31091438217071E-003
-7.12445384119129E-003  +1.16630740782393E-002  +5.54209858992282E-003
```

```
-5.08654095997511E-003   +5.49364290599985E-003   +2.47868607962678E-003
-4.42682322448696E-003   +3.39406077841307E-003   +1.59226159967363E-003
+2.64750543903317E-003   +2.48745633134807E-003   +1.05200031597483E-003
+2.56865154271922E-003   +1.68183121634717E-003   +6.24561494858841E-004
+3.03410853857312E-003   -1.11132103372786E-003   -1.26184453339161E-003
```

**Table 2.** Optimal ratios $m_{Sun}/m_{planet}$ for Mercury, ..., Pluto

```
6023615.18470910      408521.161125963     328900.711579757
3098698.66262312      1047.34870086844     3497.89111808851
22902.9182183941      19412.1792363837     134998758.004486
```

Apart from the ratios $m_{Sun}/m_{Pluto}$ and $m_{Sun}/m_{Venus}$, which decreased respectively (in relative amounts) by $1.5 \times 10^{-3}$ and $6.2 \times 10^{-6}$, the ratios $m_{Sun}/m_{planet}$ did not change by more than 2 or 3 $\times 10^{-6}$ (Mercury, Mars, Saturn, Uranus, Neptun) or less ($5 \times 10^{-7}$ for the EMB and $10^{-7}$ for Jupiter). The product $Gm_{Sun} = G$ decreased, in relative amount, by $1.8 \times 10^{-9}$, to reach $2.95912207762858 \times 10^{-4}$ au$^3$/day$^2$. In relation with the different accuracies to which these masses are supposed to be known, the most important change is the one for the EMB. It is emphasized again that the masses are model-dependent, hence they do not have to take exactly the same values with the purely Newtonian equations of motion used here, as the values they take for a calculation involving equations of motion based on GR, like DE403. (*Cf.* Arminjon [7] for a theoretical discussion, in this line, on the adjustment of masses.) Moreover, since the perturbations due to the Moon were taken into account only in an approximate way, the values found here cannot be really optimal even in the purely Newtonian framework. The neglect of all perturbations due to the asteroids also decreases the accuracy of the comparison, especially for Mars. However, all of these perturbations are Newtonian and the techniques to account for them are known (but time-consuming).

The main question that might be answered by the present calculation is whether there is some possibility that the effect of the standard corrections of GR (which is the strongest for Mercury) could be at least partly replaced by a purely Newtonian calculation using optimized parameters. Figure 1 shows, for Mercury ($i = 1$), the time evolution of the norm

$$\delta R_i(t) \equiv \|\mathbf{r}_i(t)_{calculated} - \mathbf{r}_i(t)_{DE403}\| \quad \text{(au)}, \qquad (11)$$

the value $\mathbf{r}_i(t)_{calculated}$ being calculated with the Newtonian *N*-bodies program, using either the initial set of parameters or the optimized set. The time drift shows the same figure with both sets, but it is ten times smaller with the optimized set. There has been no emphasis on Mercury in the calculation, and the improvement indeed does not especially concern Mercury. Table 3 gives the maximum over the time *t* (spanned by steps of 20 days between JD 2451600.5 and JD 2416800.5), of the norm (11), and the improvement for this error measure, as compared with the maximum error obtained with the initial set of parameters. Thus, the maximum error on the radius vector *r* of Venus is 103 times smaller with the optimal set of parameters than with the initial set.

**Table 3.** Maximum error on the radius vector (au) over 34800 days with the optimal set of parameters, followed by the improvement factor over the maximum error obtained with the initial set.

Mercury: $(3.2 \times 10^{-5}, 9.69)$.   Venus: $(5.5 \times 10^{-7}, 103)$.   EMB: $(1.2 \times 10^{-6}, 35.6)$.
Mars:    $(1.8 \times 10^{-6}, 15.3)$.   Jupiter: $(9.5 \times 10^{-8}, 43.3)$.   Saturn: $(3.3 \times 10^{-8}, 51.0)$.
Uranus:  $(1.2 \times 10^{-8}, 47.2)$.   Neptune: $(1.1 \times 10^{-8}, 64.3)$.   Pluto: $(1.2 \times 10^{-8}, 9.06)$.

The angular error on the heliocentric positions was also calculated (Table 4), according to:

$$[\delta\varphi_i(t)]_{\text{arc seconds}} = \|\boldsymbol{n}_i(t)_{\text{calculated}} - \boldsymbol{n}_i(t)_{\text{DE403}}\| \times 3600 \times 180/\pi, \quad \boldsymbol{n} \equiv \boldsymbol{r}/\|\boldsymbol{r}\|. \quad (12)$$

**Table 4**. Maximum, over 34800 days, of the angle between the heliocentric position of Mercury, ..., Pluto, according to DE403, and the position calculated with the optimal set of parameters. (Arc seconds)

```
19.5312       0.158437      0.249372      0.249408      0.00396196
0.000666968   0.000130530   7.01700e-005  8.19931e-005
```

It is a very small error, except for Mercury, for which, however, this angular error (20.5" per century) is less than the halfth of the standard residual advance in perihelion (43" per century). And the comparison with the angular error which is got with the initial set of parameters shows again an improvement by a factor ten (Fig. 2). Moreover, the same figure and the same rates are found for the angular error on Mercury's heliocentric position, if one compares, with the "long" ephemeris DE406, a calculation spanning 30 centuries (Fig. 3). Recall that the input data used in the fitting span a time interval which increases with the revolution period of the planet, and for Mercury the time interval is merely 400 days. Thus, the calculation whose result is illustrated on Fig. 3 represents an extrapolation of the time range, as compared with the range containing the input data, by a factor 2740. Yet the maximum angular error still increases merely linearly with time, even over that greatly extrapolated range. In particular, the optimized set of parameters still shows the same advantage over the initial set. Thus our optimization behaves robustly as regards extrapolation, at least for Mercury.

Hence one might expect that the longitude of the perihelion, $\varpi$, has been improved by the parameter optimization. But that is not the case: for Mercury ($i = 1$), the difference $\delta\varpi_i(t) = \varpi_i(t)_{\text{calculated}} - \varpi_i(t)_{\text{DE403}}$ shows a linear drift of just 43" per century, whether the Newtonian calculation is done with the initial set of parameters or with the optimal one (Fig. 4). (Also for this result, it has been checked that it remains unchanged if the calculation spans 30 centuries, substituting DE406 for DE403.) This could well be taken to imply that the advance in the perihelion is not an adequate measure of the angular error on a planet's trajectory. It is true that the definition of the osculating elements is somewhat abstract. And indeed, the comparison between Figs. 2 and 4 shows that the relation between the angular error and the advance in the longitude of the perihelion (the latter being calculated from the osculating elements) is very loose. However, see the remarks at the end of the paper.

*4.3 Other optimization runs*
1) In a second optimization run, an attempt was made at taking weights proportional to the inverse of the uncertainty of the data. The following weights were adopted for the nine planets (Mercury to Pluto): $w_i$ = (10, 5, 10, 10, 0.1, 0.05, 0.025, 0.015, 0.0062), and only the positions, not the velocities, were included in the input data. However, much more input data were taken from the ephemeris: 551 positions per planet (thus spanning 11000 days instead of 400, for Mercury), instead of 21. Yet, the improvement found for the angular error on the position of Mercury was only a bit more, at the cost of some significant mass modifications. It may be safer to keep the masses of the planets fixed in the optimization, except for the masses of the giant planets (Jupiter, Saturn, Uranus, Neptune). However, in the first optimization discussed, the masses modifications remained small, even for the other, much less massive planets.
2) In a third optimization, the same (unequal) weights and the same input data were adopted as in the second optimization, but the masses of the non-giant planets were kept fixed at the values of the DE403 ephemeris. Again, no significant improvement was found for the angular error on the position of Mercury, as compared with the first optimization.
3) In a fourth optimization, equal weights were adopted (as in the first one), but all masses were kept fixed at the DE403 values. This gave very slightly better, but nearly undistinguishable results from those of the first optimization (with all masses left free). This result shows that the first optimization did not reach the exact absolute minimum of the residual (since the latter minimum cannot be higher than that got with all masses fixed). More importantly, it shows that it is only the adjustment of the initial data that leads to the lowering of the global residual by a factor 24.

**6. Summary and discussion**

With current programming facilities, it becomes easy to build accurate programs for the Newtonian *N*-bodies problem and the calculation of osculating elements. One particularity of the present work is the attempt made at simultaneously optimizing all parameters of the *N*-bodies problem: constant of gravitation, masses, and initial conditions. The algorithm used for the minimization of the quadratic residual combines the linearization of the functions involved (Gauss algorithm), with the one-dimensional minimization in the descent direction given by the linearization. This converges in few iterations, but each calculation of the residual *R* involves solving the equations of motion. This must be done ($K+1+p$) times at each global iteration, where *K* is the number of parameters and *p* the (variable) number of evaluations of *R* for the one-dimensional minimization (typically $p$ = 15 to 30, due to the fact that a stringent tolerance $\eta \approx 10^{-8}$ has to be used for the latter). Hence the parameter optimization is time-consuming.

This parameter optimization was applied to check whether it is possible to reproduce, up to a small error, the predictions of an ephemeris involving corrections based on general relativity, by using a purely Newtonian calculation. With a very slight modification of the parameters, one may indeed improve the mean quadratic difference between the input data taken from the ephemeris and their values obtained by the Newtonian calculation, by a factor twenty-four. Even for Mercury, the difference is reduced by a factor ten. It has been found that the improvement in the accuracy of the angular position is not well-described by the advance in perihelion: while the angular position of Mercury was improved by a factor ten, the advance in perihelion remained virtually unaffected at the standard value of 43" per century. Even with optimized parameters, the difference between the Newtonian calculation and the relativistic ephemeris remains some 20" per century for the angular position of Mercury, which is more than two orders of magnitude above the observational accuracy, considered to be 0.1" for the period 1900-2000 [10, p. 228]. Thus, *the answer to the question at the beginning of this paragraph is definitely* « NO ». It is yet interesting to note the contrast between the improvement in the effective angular position of Mercury, and the absence of any improvement found for the longitude of the perihelion. Indeed it is the latter, indirect measure of the angular accuracy, that is most usually referred to in the discussion of the observational tests of the relativistic theories of gravitation (*e.g.* Fock [8, pp. 215-221], Weinberg [16], Will [17]). However, the improvement obtained from the optimization is essentially due to the optimization of the initial conditions, not that of the masses. Thus, the relatively high angular drift of 220"/cy with respect to DE403, as found for Mercury after a Newtonian calculation using the initial conditions directly taken from the DE403 ephemeris, reflects merely the fact that the initial conditions must be re-optimized when one changes the model. (In turn, this means that the adequate initial conditions actually depend on the precise model that is being used.) If this is systematically done, the advance of perihelion is likely to become one meaningful measure of the angular error, though not a sufficient one by it alone.

The present computer program has been also used [18] to optimize the initial conditions over a much longer time range (60 centuries), for two different models: a Newtonian model restricted to the Sun and the four giant planets, and a Newtonian model corrected by the Schwarzschild effects of the Sun, and including the nine major planets.

**Acknowledgement.** I am grateful to P. Bretagnon, Cl. Le Guyader and C. Marchal for helpful remarks during the course of this work. The remarks of E.M. Standish on the manuscript led me to make several improvements in its redaction and to make the additional calculation of Fig. 3. The data of the DE403 and DE406 ephemerides were read from the web site of the Institut de Mécanique Céleste du Bureau des Longitudes, Paris.

# Figure captions

Fig. 1. Norm of the difference in the heliocentric position vectors, as obtained either from the relativistic ephemeris DE403 or by a Newtonian calculation, and this one either with the initial set or the optimal set of parameters.

Fig. 2. Angular difference in the heliocentric position vectors, as obtained either from the relativistic ephemeris DE403 or by a Newtonian calculation, and this one either with the initial set or the optimal set of parameters.

Fig. 3. As for Fig. 2, but here for a long time range (30 centuries instead of 95.3 years), with DE406 instead of DE403.

Fig. 4. Difference in the longitude of perihelion, as obtained either from the relativistic ephemeris DE403 or by a Newtonian calculation, and this one either with the initial set or the optimal set of parameters. (The two calculations are hardly distinguishable.)

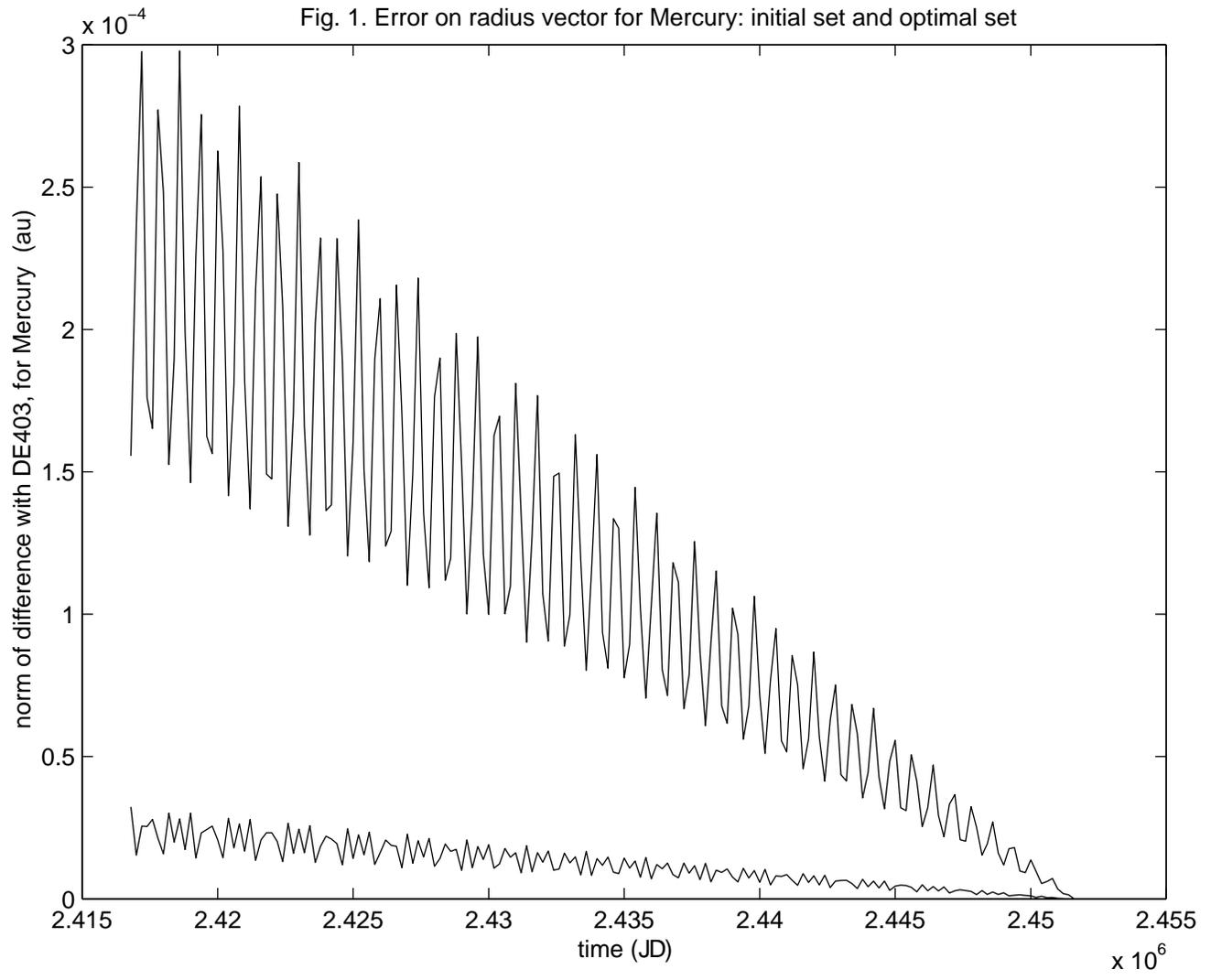
Fig. 1. Error on radius vector for Mercury: initial set and optimal set

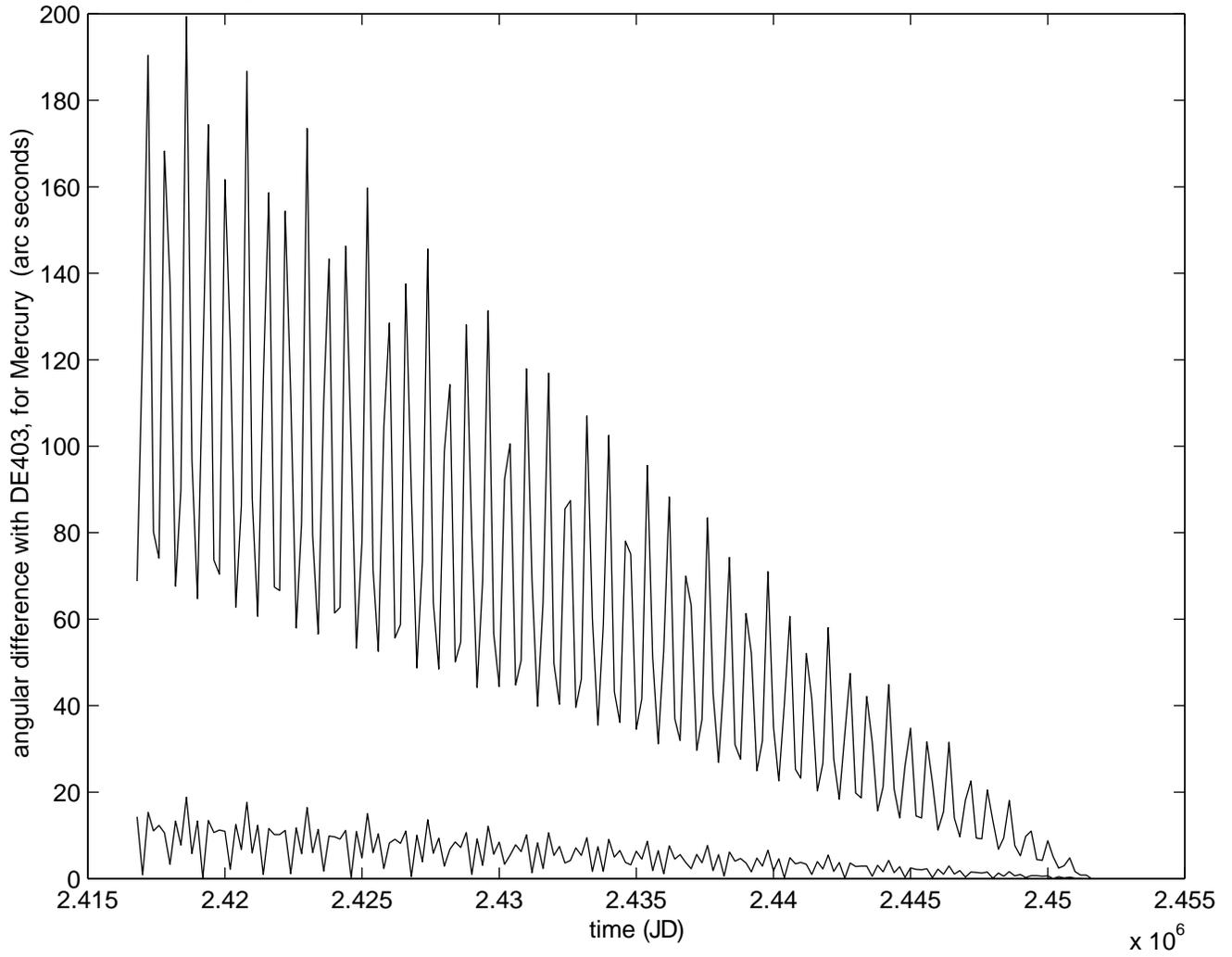

Fig. 2. Angular error on heliocentric positions for Mercury: initial set and optimal set

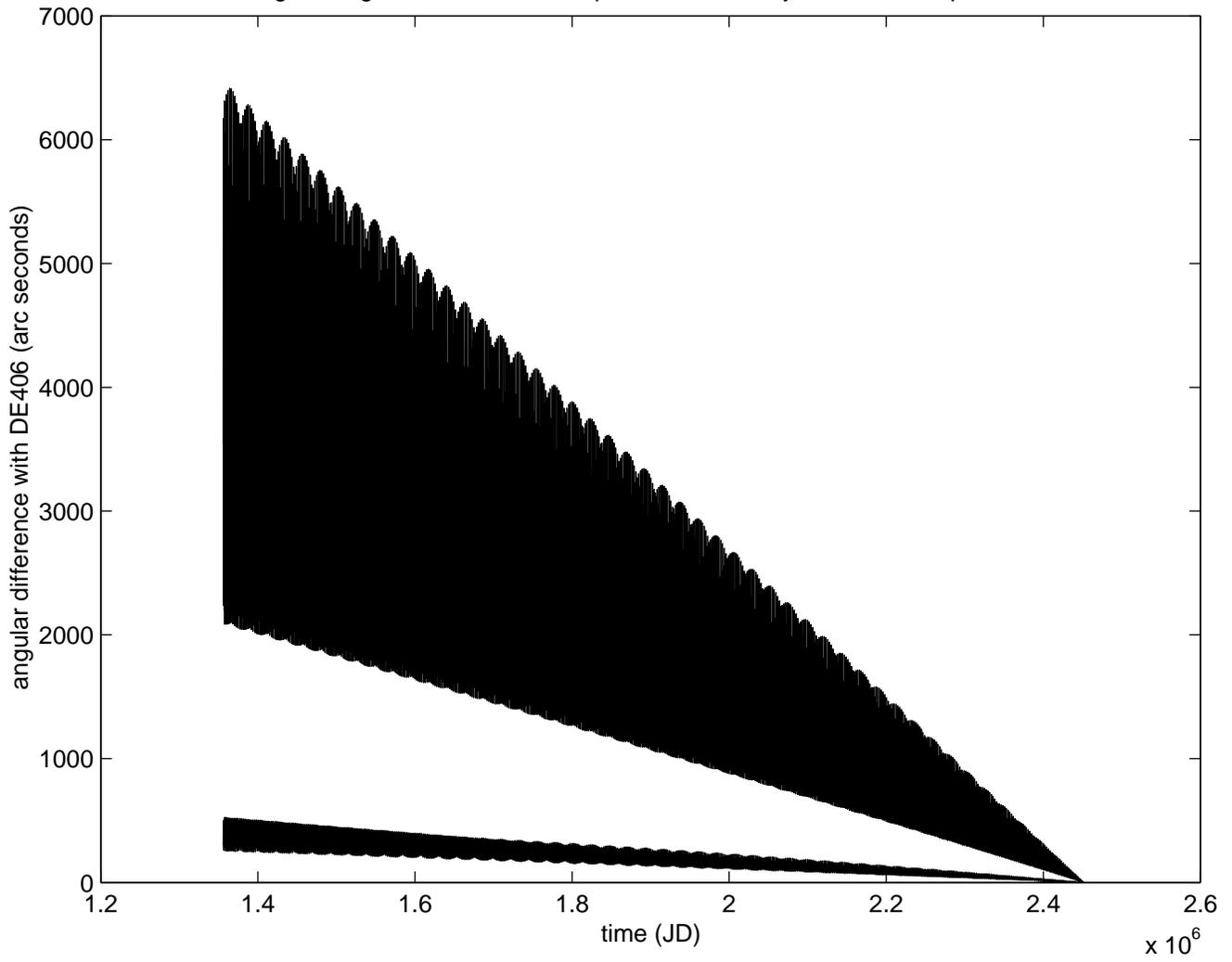

Fig. 3. Angular error on helioc. positions, Mercury: initial set & optimal set

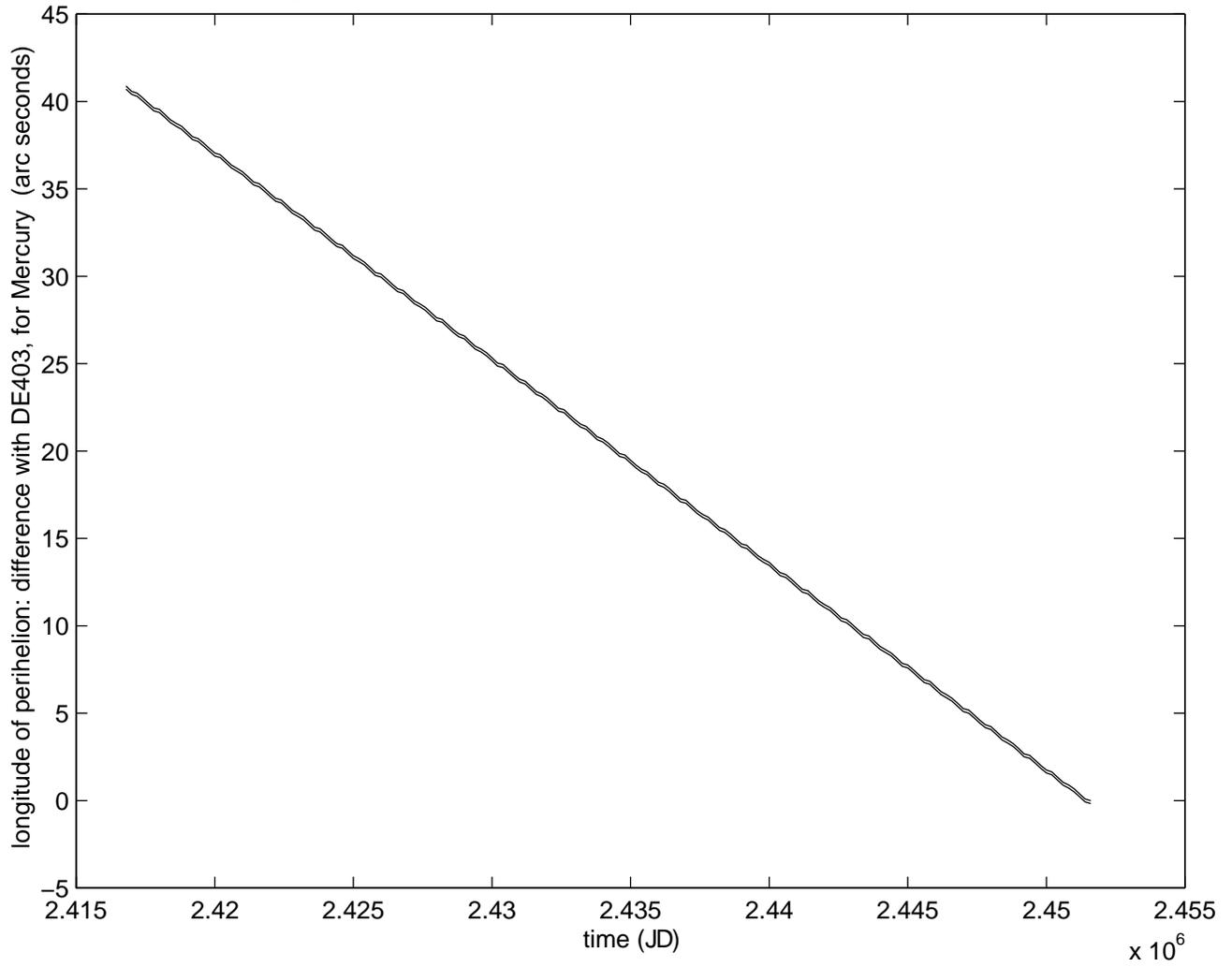

Fig. 4. Error on perihelion position for Mercury: initial set and optimal set